# Low-loss single-mode hybrid-lattice hollow-core photonic crystal fiber

Foued Amrani,[1, 2] Jonas H. Osório,[1] Frédéric Delahaye,[1, 2] Fabio Giovanardi,[3] Luca Vincetti,[3] Benoît Debord,[1, 2] Frédéric Gérôme,[1, 2] Fetah Benabid[1, 2]

[1]GPPMM Group, XLIM Institute, CNRS UMR 7252, University of Limoges, Limoges, 87060, France
[2]GLOphotonics, 123 Avenue Albert Thomas, Limoges, 87060, France
[3]Departament of Engineering "Enzo Ferrari", University of Modena and Reggio Emilia, Modena, 41125, Italy
* f.benabid@xlim.fr

**Abstract:** The remarkable recent demonstrations in ultralow loss Inhibited-Coupling (IC) hollow-core photonic crystal fibers (HCPCFs) place them as serious candidates for the next-generation of long-haul fiber optics systems. A hindrance to this prospect, but also to short-haul applications such as micromachining, where stable and high-quality beam delivery is needed, is the challenge to design and fabricate an IC-guiding fiber that combines ultra-low loss, truly and robust single-modeness, and polarization-maintaining operation. Design solutions proposed up to now require a trade-off between low loss and truly single modeness. Here, we propose a novel concept of IC HCPCF for obtaining low-loss and effective single-mode operation. The fiber is endowed with a hybrid cladding composed of a Kagome-tubular lattice (HKT). This new concept of microstructured cladding allows to significantly reduce confinement loss and, at the same time, preserving a truly and robust single-mode operation. Experimental results show a HKT-IC-HCPCF with a minimum loss figure of 1.6 dB/km at 1050 nm and a higher-order modes extinction ratio as high as 47.0 dB for a 10 m long fiber. The robustness of the fiber single-modeness was tested by moving the fiber and varying the coupling conditions. The design proposed herein opens a new route for the accomplishment of HCPCFs that combine robust ultralow loss transmission and single-mode beam delivery and provides new insight into the understanding of IC guidance.

## 1. Introduction

Due to their excellent performance as a platform for the study of fundamental physics and for addressing applied problems in photonics, hollow-core photonic-crystal fibers (HCPCFs) continue to be the subject of intense research since its theoretical proposal in 1995[1]. Such interests are driven by the accomplishment of outstanding results in science, such as atom optics[2] and gas-based nonlinear optics[3], and in industry, such as ultra-short-pulse and high energy laser beam delivery[4] to mention a few. Furthermore, the physics of the HCPCF optical guidance is still a fascinating and ongoing topic of research[5].

Today, we distinguish two types of HCPCF; one type that guides via photonic bandgap (PBG)[1], and the second type that guides via Inhibited-Coupling (IC)[6] mechanism. In PBG fibers, light is guided because the fiber microstructure is such that there is no cladding mode to which the core mode can couple to. Conversely, in IC fibers, although there is no bandgap in the cladding frequency-effective refractive index space, the coupling of the core mode to the cladding is strongly inhibited due to a low spatial overlap between the core and cladding mode field's distribution, and to a strong mismatch between these modes transverse spatial phases[7]. These principles provided the conceptual tools to introduce the hypocycloid core contour (*i.e.*, negative curvature) concept[8, 9], which enabled dramatic enhancement in light confinement of such fibers, as exemplified by IC guiding hypocycloid core-contour Kagome-lattice HCPCF, and single-ring tubular-lattice (SR-TL) HCPCF[10]. Experimental illustrations of the negative curvature core-contour impact are the reduction of the loss figures in Kagome HC-PCFs down to 8.5 dB/km at 1030 nm[11], which is considerably lower than the dB/m loss level reported in the first Kagome fiber[12], and the realization of optimized SR-TL HCPCFs with transmission loss as low as 7.7 dB/km at 780 nm[7] and 13.8 dB/km at 539 nm[13]. Among the noteworthy conclusions of such effort is the fact that, for wavelengths shorter than ~1 µm, the transmission loss is no longer limited by the cladding design. Instead, the surface roughness-induced scattering loss (SSL) is the limiting factor. On the other hand, for wavelengths longer than ~1 µm, the confinement loss (CL), and hence the cladding design, remains the limiting factor of the IC HCPCFs transmission performance[7]. Very recently, new negative curvature cladding designs were introduced, and lower CL than Kagome-HCPCF and SR-TL HCPCF have been demonstrated. Very promising transmission loss figures were reported, as exemplified by the figures of 2 dB/km at 1512 nm for conjoined tubular cladding HC-PCF[14], and 0.28 dB/km at 1550 nm for nested[15] tubular cladding HC-PCF.

While the recent decreasing trend in IC HCPCFs loss figures is remarkable, the current challenge in the field is to design and fabricate an IC-guiding fiber that combines ultra-low loss, truly single-mode, and polarization-maintaining operation, especially at wavelengths much shorter than 1550 nm, such as around 1 µm. To illustrate the difficulty of only combining single-mode and ultra-low loss in IC-guiding fibers at ~1 µm, we recall the following. Rigorously speaking, in IC-guiding fibers, single-modeness is impossible because of the ubiquitous presence of higher-order modes in the fiber core. However, one can get closer to a truly single-mode operation if the loss extinction ratio between the lowest loss mode (typically, the core fundamental mode) and the second-lowest loss mode is sufficiently high. It was successfully achieved in six-tube SR-TL HCPCF thanks to the adequate ratio between the core and lattice tubes diameters, $D_{tubes}/D_{core}$, calculated to be 0.68, which provides effective refractive index matching between the $LP_{11}$-like modes of the core (which are usually the main contaminating higher-order mode in the fiber modal content) and the fundamental, $LP_{01}$-like, mode of the lattice tubes[16-18]. The inconvenience of this approach, however, is that the CL of the core fundamental mode is considerably high because of the small core diameter. For example, Gao *et al.* measured loss values around 500 dB/km loss for wavelengths around 1 μm with six-tube SR-TL HCPCFs[19]. On the other hand, the minimum loss figure reported in the literature using IC-guiding fibers around 1 μm is 2.5 dB/km, which was obtained by using a nested tubular cladding HCPCF[20]. However, the ratio between the loss values of the fundamental and $LP_{11}$-like modes in nested fibers is typically lower than 20 dB[15], which is considerably inferior to the ratio achieved in six tubes-lattice SR-TL HCPCF, which can be higher than 30 dB[18]. Whilst such a ratio is sufficient to have an effective single-mode operation under static conditions by suitable input light coupling, it becomes problematic to ensure single-modeness in conditions where the fiber is under constant motion and bend.

Here, we present the design and the fabrication of a novel HCPCF structure that combines single-mode operation and ultra-low CL. The cladding design exhibits a hybrid lattice made of Kagome and tubular cladding lattices. The utilization of two IC claddings allows to decrease CL while keeping the fiber core demarcated by a six tubes-tubular lattice for effective single-mode operation. As it will be shown in the following, a minimum loss figure of 1.6 dB/km at 1050 nm was experimentally achieved by using this fiber design. Moreover, the modal content of the fiber was measured by using spatially and spectrally ($S^2$) imaging technique[21]. The higher-order modes' contributions were measured to have a maximum extinction ratio as low as 47.0 dB for a 10 m long fiber at optimized coupling conditions. Finally, the robustness of the fiber single mode character was verified by inspecting the output mode while the fiber was moving and by changing the coupling conditions. We believe that the fiber design proposed herein provides a new avenue for obtaining low-loss single-mode HCPCFs.

## 2. Results

### 2.1 Design rationale

Fig. 1 summarizes the proposed fiber structure, the hybrid Kagome-tubular lattice (HKT) HCPCF, and its design rationale. Fig. 1a shows the fiber transverse geometrical microstructure. It is endowed with a two lattice cladding. The inner cladding consists of a lattice of six tubes, which demarks the fiber core. The six-tube lattice is chosen for attaining effective single-mode operation via resonant filtering of the $LP_{11}$ mode[16-18]. Also, this inner cladding stands out with the absence of connection nodes, which favors the IC guidance[7]. The outer cladding of the HKT-HCPCF comprises a Kagome lattice structure in which the tubular lattice is embedded. As it will be demonstrated in the following, the association of two IC claddings enhances the light confinement in the core significantly and reduces the fiber CL expressively. Fig. 1b shows the cross-sections of the fiber designs (FD) we study herein (*lhs*) and their respective simulated CL as a function of normalized frequency values, $F = (2t/\lambda)\sqrt{n_g^2 - 1}$ (*rhs*). Here, $t$ and $n_g$ represent the thickness and the refractive index of the cladding glass, respectively; $\lambda$ is the wavelength. In CL calculation, all the fibers were taken to have the same core inner-diameter (33.5 μm), $t$ (1100 nm) and $n_g$ (1.45). The first design (FD #1, purple line) consists of a jacketless 6-tube lattice structure whose minimum CL values are around 1650 dB/km, 20 dB/km, and 3 dB/km in the fundamental, first and second-order transmission bands respectively. The second design (FD #2, green line) reproduces a 6-tube lattice SR-TL HCPCF. The results show that jacketing the suspended tubes impacts the CL spectrum mildly with a rise of the above loss figures to around 1900 dB/km, 40 dB/km, and 5.5 dB/km, respectively. The blue line in Fig. 1b shows the CL for a Kagome-lattice (KL) HCPCF, identified as FD #3. It is seen that, for this design, the minimum CL figures are around 440 dB/km, 50 dB/km, and 30 dB/km in the fundamental, first and second-order transmission bands. We note for this fiber the oscillating structure in the CL spectrum due to connecting struts or nodes[6], which can either locally increase or decrease the CL compared to FD#1 and FD#2.

The HKT-HCPCF we propose herein is identified in Fig. 1b as FD #4. The red line in Fig. 1b shows the ideal case of such a structure. In this situation, there is no physical connection between the tubular and the Kagome lattices, and the spacing between the lattices is set to be 1.59 μm. Although this is not a feasible fiber design, it communicates the potential of this novel design to achieve impressive low CL values – as low as 0.35 dB/km, 7 × 10$^{-5}$ dB/km and 8.6 × 10$^{-6}$ dB/km for wavelengths within the fundamental, first and second-order transmission bands respectively. It is noteworthy that the two latter values for this ideal structure are well below the current attenuation level of solid-core fibers, represented as a grey dashed line in Fig. 1b[22]. Also, there is an expressive difference of five orders of magnitude between the CL of FD #4 and the FD #1, FD #2, and FD #3 ones, as emphasized by the black arrow in Fig. 1b. Fig. 1c shows the evolution with the spacing between the Kagome and tubular lattices (*g*, see inset of Fig.1c) of the minimum CL for the 1$^{st}$ (blue curve) and 2$^{nd}$ (red curve) transmission bands. The results readily show that, for a significant drop in CL, *g* must be larger than a critical value $g_{cr}$. When *g* is varied from 0.75 μm to 5.85 μm, CL drops from 5.11 × 10$^{-3}$ dB/km to 2.40 × 10$^{-4}$ dB/km at F = 1.66, and from 3.69 × 10$^{-5}$ dB/km to 2.79 × 10$^{-6}$ dB/km at F = 2.58. The case when *g* = 0 is also shown in Fig. 1c. For this one, it is seen that the CL values are much higher, with values as high as 2.95 dB/km at F = 1.66, and 0.30 dB/km at F = 2.58. In these simulations, the alteration in the $g$ values was obtained by adequately enlarging the Kagome lattice pitch so the core diameter ($D_{core}$ = 33.5 μm), the distance between the cladding tubes ($\delta$ = 4.67 μm), and the tubes and Kagome lattices thickness ($t_{tubes} = t_{kago}$ = 1100 nm) could be maintained. Based on the data presented in Fig. 1c, the critical $g$ value ($g_{cr}$) was found to be around 1.5 μm for attaining the dramatic reduction of the CL in the HKT-HCPCF design comparatively to usual Kagome and tubular fiber designs. It is noteworthy that the results of our numerical study (not shown) also informs that the azimuthal position of the inner cladding relative to the outer cladding also affects the CL.

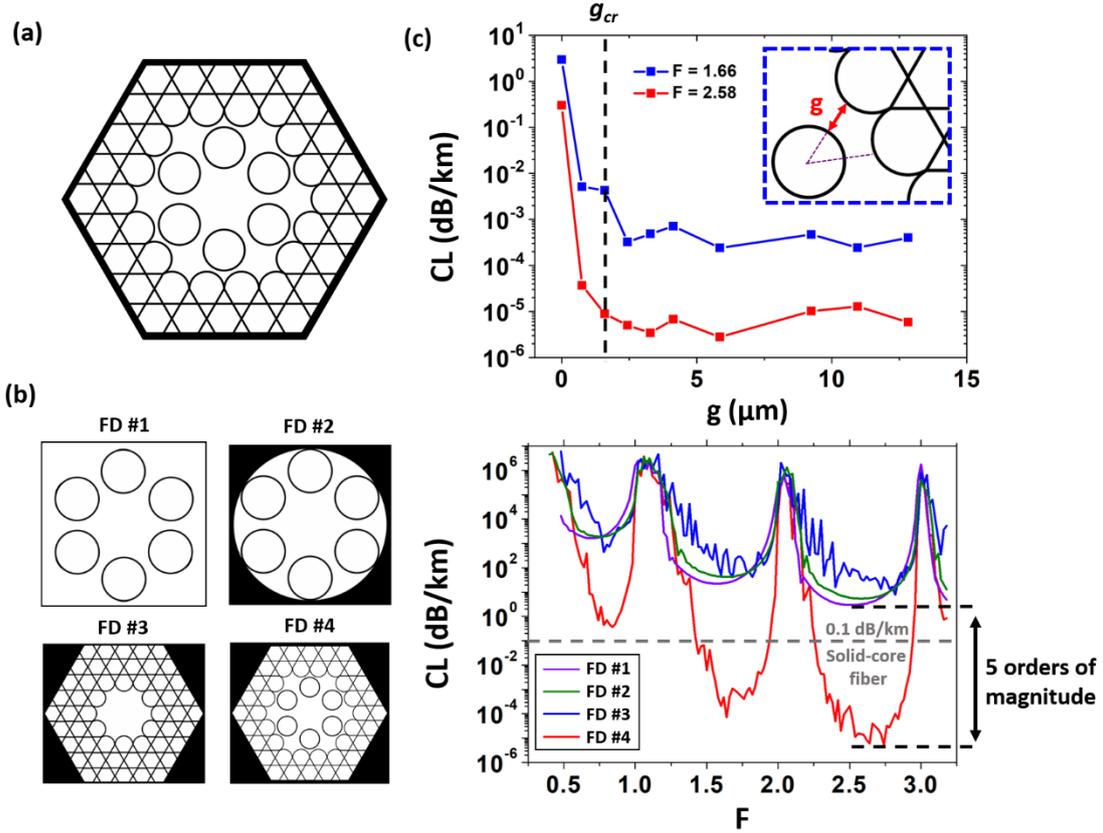

**Fig. 1.** (a) Schematic diagram of the hybrid Kagome-tubular HCPCF. (b) CL simulation results for different fiber designs (FD). FD #1: jacketless tubular lattice; FD #2: jacketed tubular lattice; FD #3: Kagome lattice; FD #4: hybrid lattice. (c) CL values as a function of the distance between the tubular and Kagome lattices ($g$) for selected normalized frequency values (F).

In order to give a physical insight into the increase of the confinement power in the HKT lattice cladding, we recall that, in contrast with PBG fibers, the CL in Kagome fibers does not monotonically decrease with an increasing number of cladding layers. Instead, the minimum CL in these fibers is reached for an optimum number of layers, which stems from a trade-off between the cladding structure confining power and the growth of the cladding density of photonic states[6]. Indeed, although adding cladding layers improves the confinement of the mode in the core, it concurrently creates additional cladding modes to which the core mode is weakly coupled to, entailing thus further CL. Therefore, the optimum CL in Kagome lattices is accomplished by adequately considering the compromise between the number of photonic states in the cladding and the confining power of the structure. This feature is investigated and corroborated in a systematic study summarized in Fig. 2 and Fig. 3.

Fig. 2 shows the CL by considering a 6-tube lattice structure and sequentially adding the Kagome structure around the tubular cladding layer by layer. For simplicity, we define a parameter $\xi$, which stands for the ratio between the thickness of the considered outer cladding in the simulations and the Kagome cladding pitch. In our analysis, $\xi$ is varied from 0 (no Kagome cladding) to 2 (Kagome lattice composed of two rings of tubes), and the calculated CL values are presented in Fig. 2a. Fig. 2b shows the CL values for two representative normalized frequency values (F = 1.66 and F = 2.58). It is seen that the effect of adding the Kagome cladding on the CL drop rate is extremely drastic between $\xi = 0$ and $\xi = 0.5$, and gentler for $\xi > 1$. At F = 1.66, the CL is calculated to be 17.4 dB/km when $\xi = 0$ and $6.5 \times 10^{-3}$ dB/km when $\xi = 0.5$. At F = 2.58, the CL was accounted as 1.5 dB/km when $\xi = 0$ and $1.6 \times 10^{-4}$ dB/km when $\xi = 0.5$. When $\xi = 1$ and $\xi = 2$, a further decrease, but with a slower rate, is observed in the CL values. At F = 1.66, the CL is calculated as $1.1 \times 10^{-3}$ dB/km when $\xi = 2$. At F = 2.58, CL is calculated as $1.0 \times 10^{-5}$ dB/km when $\xi = 2$. Here, it is worth saying that having a Kagome cladding formed by more than two rings of tubes (i.e., having $\xi > 2$) entails marginal effect in decreasing the CL[23]. In the simulations shown in Fig. 2, the core diameter ($D_{core}$ = 33.5 μm), the distance between the cladding tubes (δ = 4.67 μm), the distance between the tubular and kagome lattices (g = 1.59 μm), and the tubes and Kagome lattices thickness ($t_{tubes} = t_{kago}$ = 1100 nm) were kept constant.

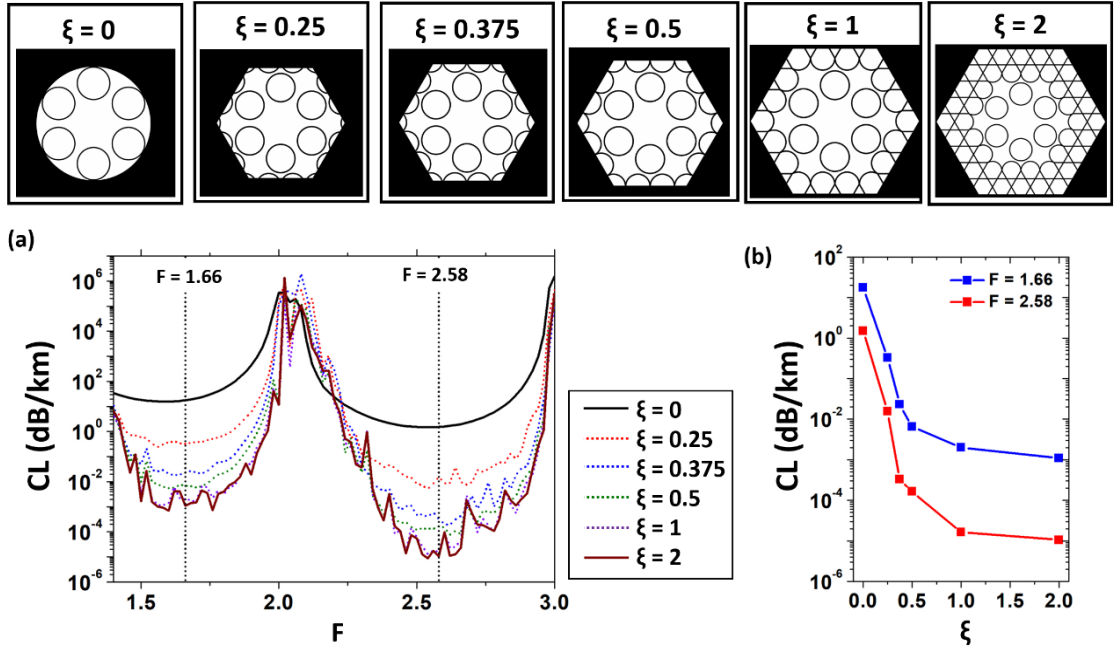

**Fig. 2.** Study on the effect of adding the Kagome lattice around the tubular lattice in the HKT-HCPCF design. (a) CL simulation results for fiber designs with different ξ values (see text for the definition). (b) CL values for two selected normalized frequency (F) values.

Fig. 3 presents the transverse energy flow for the different ξ. It shows the diagrams of the radial component of the Poynting vector, which proved to be a useful tool in examining the leakage dynamics in IC fibers[7]. In the figure, we plot the normalized component of this quantity, $\bar{\bar{p}}_r = \frac{1}{2p_z} \vec{E} \times \vec{H}^* \cdot \hat{r}$ (where $\vec{E}$ and $\vec{H}$ are the electric and magnetic fields, $\hat{r}$ is the radial unit vector, and $p_z$ is the maximum value of the longitudinal component of the Poynting vector), at F = 2.58 for representative values of ξ. Consistently with the results published by Debord et al.[7], the distance between the cladding tubes in the fiber design studied here (δ = 4.67 μm) entails the main channel for the core fundamental mode leakage to be the direction through the lattice tubes (instead of the direction through the gap between the cladding tubes; see the results for ξ = 0). For ξ > 0, one sees that the presence of the Kagome lattice around the tubular one minimizes the power flux through this leaking channel of the tubular lattice structure. This leakage reduction increase with increasing ξ for the values considered here. It is noteworthy that the main leaking channels for the Kagome lattice are azimuthally shifted from the direction through the inner lattice tubes.

Fig. 4 situates the HKT lattice design potential for offering excellent CL and higher-order modes (HOM) extinction levels among representative IC HCPCFs designs explored in the literature. Fig. 4a shows a comparative plot between the CL of the fundamental mode in SR-TL HCPCF (I, green line), nested (II, pink dashed line), straight-bar (III, orange dashed line), nested-rod (IV, purple dashed line) and conjoined tubes (V, gray dashed line), and the hybrid design proposed herein (VI). The data stands for fibers with a core diameter of 30.5 μm, and lattice tubes with outer diameter and thickness of 22.1 μm and 1.1 μm, respectively. In design V, the tubes in the second ring layer have a diameter of 26.4 μm, and, in the designs II and IV, the nested tubes have a diameter of 12.17 μm. These parameters were considered to allow direct comparison to the data recently published by Habib et al.[24].

In Fig. 4a, it is seen that the fiber designs II, III, IV, and V allows decreasing the CL values of SR-TL HCPCFs (I) by around two orders of magnitude. However, their CL remains within the range of ~1-3 dB/km, with the nested design (II) having the minimum value of 0.5 dB/km for the fiber parameters considered here. In contrast, the ideal hybrid fiber structure (VI, red line) offers a dramatic CL reduction. CL values drop to 7 × 10$^{-5}$ dB/km around 1410 nm. In this simulation, the distance between the tubular and Kagome lattices was set as 2.04 μm.

While the ideal HKT-HCPCF design has suspended inner cladding tubes (and, therefore, it is not feasible), it is a structure of important academic interest for demonstrating the concept of the association of two IC claddings towards the reduction of the CL, as well as the potential of the HKT-HCPCF to provide impressive low loss figures. Taking it into account, we show in Fig. 4 (VII, blue line) a realizable version of the HKT-HCPCF. In this fiber design, the cladding tubes are connected to the Kagome lattice via the utilization of thin tubes. Indeed, simulations show that the thinner the connection tubes, the lower the CL values will be. Here, we have chosen connecting tubes with a thickness of 640 nm (i.e., 58% of the cladding struts and tubes thickness) as a reasonable value considering the HKT-HCPCF fabrication.

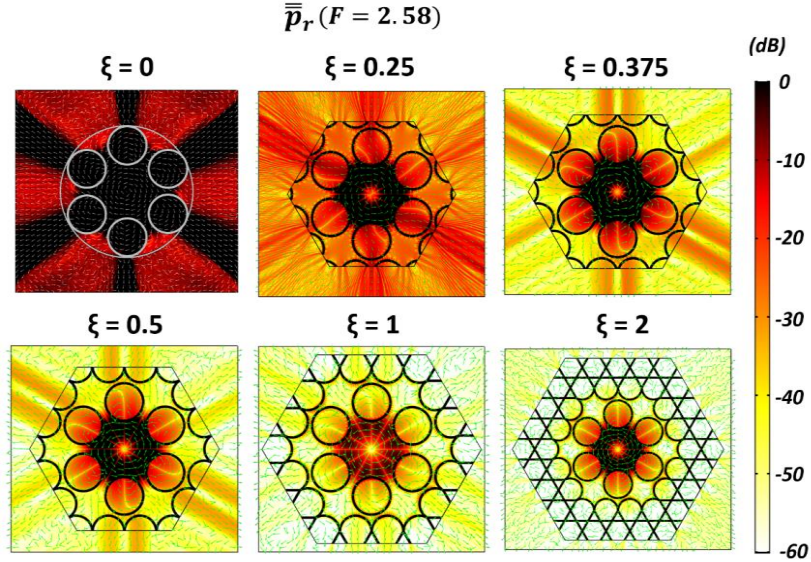

**Fig. 3.** Diagram of the normalized radial component of the Poynting vector for representative values of ξ at F = 2.58.

The results show that whilst the addition of the connecting tubes causes CL to increase, the minimum CL of $1.46 \times 10^{-2}$ dB/km around 1340 nm is 36 times lower than the lowest CL figure achieved with the other fiber designs between 1260 nm and 1400 nm – as emphasized by the black arrow in Fig. 4a.

Additionally, we evaluate in Fig. 4b the ratio between the CL of $LP_{11}$-like modes and the CL of the fundamental mode ($\alpha_{LP11}/\alpha_{LP01}$) to assess the potential of the fiber designs for single-mode operation. We distinguish two designs groups: the first one (I, V, VI, and VII) with $\alpha_{LP11}/\alpha_{LP01}$ in the range of $3\times10^2$ - $3\times10^4$, and the second group (II, III, IV) with a much lower ratio, in the range of 2-10. This difference is readily explained by the fiber design $LP_{11}$ mode resonant filtering capability[16-18]. Here, the hybrid fiber design is the only one that attains a better compromise in the binomial CL and HOM extinction. Considering the results for the hybrid fiber design, if one assumes a situation in which 99% of the power is coupled to the fundamental mode and 1% is coupled to the $LP_{11}$ mode at the fiber input, a higher-order mode extinction ratio higher than 40 dB can be estimated for a 10 m long fiber.

## 2.2 Fiber fabrication and loss measurement

Due to the potential of the HKT-HCPCF shown by the simulations, we endeavored to obtain such a fiber experimentally. The cross-section of the fabricated fiber is presented in Fig. 5a, which was obtained by using the stack-and-draw technique. The tubes that form the tubular lattice have a thickness of 1.27 µm and an inner diameter of 23.0 µm. The fiber core has 37.1 µm diameter. During fabrication, the tubes and core sizes were optimized to achieve $D_{tubes}/D_{core}$ = 0.62. This value is close to the optimum value of $D_{tubes}/D_{core}$ = 0.68, which is calculated to be the one that provides optimum coupling between the $LP_{11}$ mode guided in the core and the fundamental mode guided in the lattice tubes[18]. The Kagome lattice struts have a uniform thickness of 720 nm, and the supporting tubes have a thickness of 370 nm, both within a 50 nm variation (see Fig. 5b for a zoom in the supporting tube region).

Fig. 5c presents the loss of the HKT-HCPCF (blue curve), which was measured from a cutback measurement using 120 m and 4 m long fiber pieces. A minimum loss value of 1.6 ± 0.4 dB/km was measured at 1050 nm. Also, the simulated CL and TL (total loss) of the fundamental mode (green dotted curve for the CL and red dotted curve for the TL), and the TL of representative higher-order modes (namely, the $LP_{11}$-like and $LP_{21}$-like modes – purple and pink dotted lines, respectively) are shown in Fig. 5c. Here, it is worth clarifying that the CL values were calculated by considering the real fiber cross-section. In turn, the TL was calculated by using $TL = CL + SSL$, where SSL is the surface scattering loss, which was estimated by using $SSL = \eta F_{cc} \left(\frac{\lambda_0}{\lambda}\right)^3$, where $\eta$ is a constant linked to the surface roughness height, $F_{cc}$ is the core mode overlap with the core contour, and $\lambda_0$ is a constant which allows calibrating the SSL formula[25] (in Fig. 5c, one used $\eta = 0.25 \times 10^{-2}$ and $\lambda_0 = 1700$ nm). Good resemblance is seen between the simulated TL of the fundamental mode and experimentally measured loss. The HOM loss is found to be around two orders of magnitude higher than the loss of the fundamental mode, in good agreement with the predicted ones (see Fig. 4b).

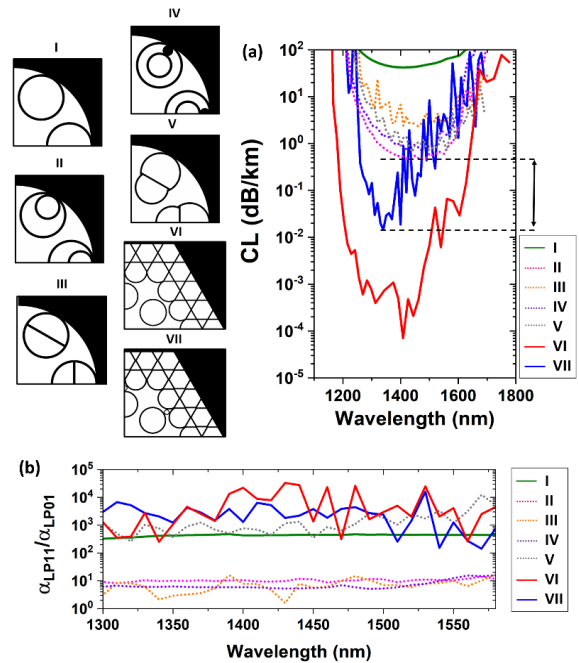

**Fig. 4.** (a) Fundamental mode CL simulation results for (I) tubular, (II) nested tubular, (III) straight-bar tubular, (IV) nested-rod tubular, (V) conjoined tubes, (VI) ideal hybrid and (VII) hybrid with supporting tubes fiber designs. (b) The ratio between the $LP_{11}$ mode and fundamental mode CL values ($\alpha_{LP11}/\alpha_{LP01}$) in the studied fiber designs.

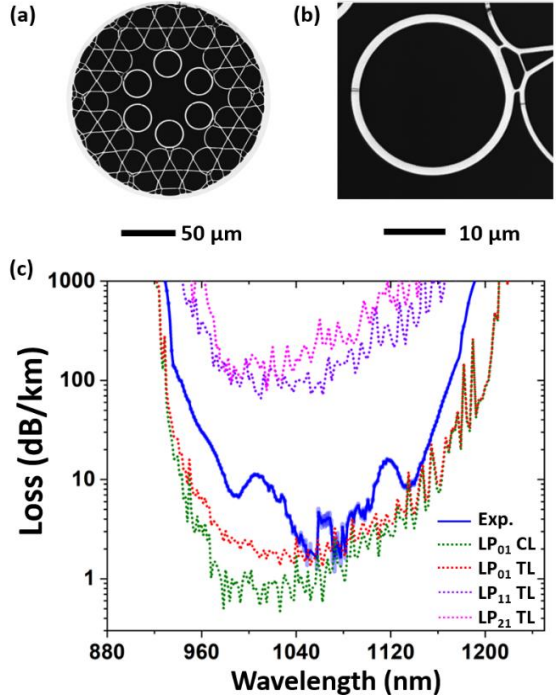

**Fig. 5.** (a) HKT-HCPCF cross-section and (b) zoom in the supporting tube region. (c) Cutback measurement results (blue curve), simulated CL and TL for the fundamental mode (green and red dotted lines, respectively), and the simulated TL for representative higher-order modes (purple and pink dotted lines).

## 2.2 Modal content measurement and further characterization results

$S^2$ measurements[21] were performed to account for the modal content of the fiber quantitatively. Fig. 6a shows the $S^2$ trace for a 10 m long HKT-HCPCF. The latter shows that no contribution of the $LP_{11}$ mode was detected (which demonstrates that the six-tube tubular-lattice fiber design has filtered it). A contribution of an $LP_{21}$-like mode was detected with MPI (multi-path interference) value as low as -47.0 dB. Table 1 shows the HOM suppression values accounted from $S^2$ measurements in IC HCPCFs. It is seen that the values reported for tubular, nested tubes, and conjoined tubes designs range from 27 dB to 22.4 dB for fibers with lengths between 10 m and 15 m. The HKT-HCPCF reported herein, therefore, allows obtaining improved effective single-mode operation in IC HCPCFs.

To show the fiber single-mode operation robustness, we conducted $S^2$ measurements by applying an offset in the input fiber position relative to the laser beam (for a fiber with length 10 m). The MPI values for the $LP_{21}$ mode are plotted in Fig. 6b as a function of the input fiber offset (the zero offset stands for the situation of optimum light transmission). It is observed that, as the coupling efficiency to the $LP_{21}$ mode grows as the input fiber offset becomes larger, the MPI values for the $LP_{21}$ mode increases from -47.0 dB to -24.7 dB when the input fiber offset is increased from 0 to 10 μm. It is noteworthy that even for such a large offset in the fiber input position, the higher-order modes contributions have a maximum MPI value of -24.7 dB in the proposed HKT-HCPCF, and that no $LP_{11}$ contribution was observed. Furthermore, Fig. 6c presents a typical $M^2$ measurement (performed at 1030 nm, using ISO standard, Metrolux LPM 200) for the HKT-HCPCF. The $M^2$ values were measured to be 1.01 and 1.02 for the x- and y-axes, respectively. Moreover, the fiber PER was tested. A maximum value of 21 dB was measured for a 10m-long fiber at 1030 nm. Also, the fiber bend loss was characterized. The measurements showed that, at 1100 nm, the bend loss values are around 0.06 dB/turn when the curvature radius is 20 cm.

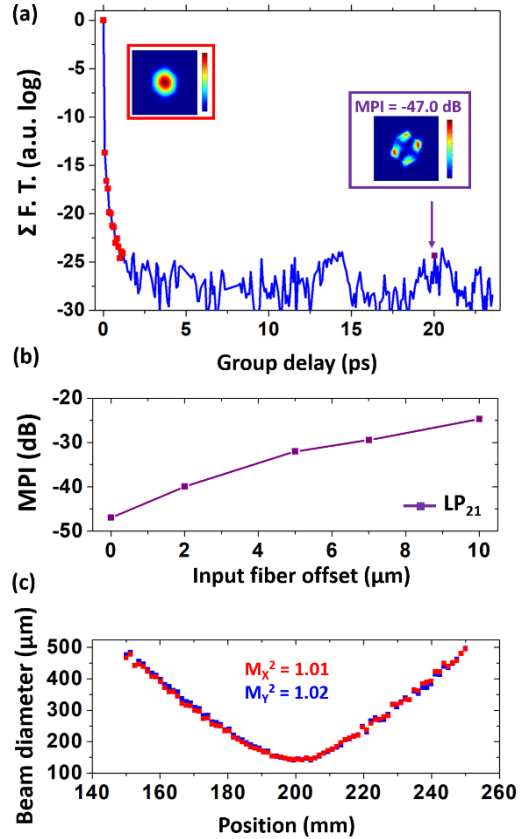

**Fig. 6.** (a) $S^2$ measurement results for a 10 m long fiber. (b) MPI as a function of the input fiber offset. (c) $M^2$ measurement results.

**Table 1.** High order mode (HOM) suppression values of IC HCPCFs measured in $S^2$ measurements.

| Fiber design | HOM suppression and fiber length | Reference |
| --- | --- | --- |
| Tubular lattice (8 tubes) | 22.4 dB for 15 m | [7] |
| Conjoined tubes lattice | > 27 dB for 15 m | [14] |
| Nested tubes lattice | ~ 25 dB for 10 m | [26] |
| Hybrid kagome-tubular lattice | 47.0 dB for 10 m | This work |

Finally, the robustness of the fiber single-mode operation against laser beam misalignment with the fiber core was tested by examining the reconstructed near field profile under different coupling conditions. To do this, we used a laser at 1064 nm and measured the power and the output near field profile as the input fiber position was scanned along the horizontal and vertical axes. The scan spans 10 μm along both axes, i.e. $\Delta x = \Delta y = \pm 5$ μm. The origin corresponds to output power maximum, measured to be 90 % of the input power. For each displacement, the fiber output power and near-field profile were recorded. Fig. 7a shows the color map of the normalized power distribution, and selected near field profiles are shown as insets (along the horizontal axis, along the vertical axis and at extreme positions). It is seen that over the 10×10μm$^2$, the near-field profile remained fundamental-like, and the transmitted power remained higher than 60 % of the maximum. It is worth saying that the color map is not perfectly symmetric due to imperfect light coupling at the fiber input.

As additional indicators of the fiber effective single-mode operation, we investigate in Fig. 7 the effect of the HOM contribution to the full width at half maximum (FWHM) and the centroid position of the output beam. Fig. 7b shows the calculated intensity mode profiles, dominated by the fundamental mode, when assuming different intensity contributions from the $LP_{11}$ and $LP_{21}$ modes to the modal content (namely 40 dB, 30 dB, and 20 dB). The plots in Fig. 7b present the intensity profiles along the horizontal axis for the cases mentioned above (the shaded areas stand for the intensity profile of the fundamental mode). It is seen that the most pronounced effect of adding the $LP_{11}$ mode to the fiber modal content is shifting the beam centroid. On the other hand, the most noticeable effect of including the $LP_{21}$ mode on the fiber modal content is changing the beam FWHM. These observations can be verified in Fig. 7c, where the normalized FWHM (FWHM/FHWM$_0$, where FWHM$_0$ is the full width at half maximum of the fundamental mode) and the centroid shift normalized by the FHWM$_0$ ($\Delta d$/FHWM$_0$) of the beams were plotted as a function of the HOM contribution to the modal content. The results show that the $LP_{11}$ and $LP_{21}$ contribution corresponding to a power extinction of 30 dB leads to a ~5% relative change in FWHM and centroid position. Fig. 7d presents graphs of the normalized FWHM and centroid variation of the output beam as the HKT-HCPCF input position was scanned along the X-axis (Y = 0 in Fig. 7a). The results show the measured FWHM/ FHWM$_0$ and $\Delta d$/FHWM$_0$ to be less than 0.5% and less than 0.2%, respectively, over the whole scan range. Also, the figure shows the HOM contribution to FWHM/FHWM$_0$ and $\Delta d$/FHWM$_0$ variation for different $LP_{11}$ and $LP_{21}$ modal combinations (dashed horizontal lines). The results show that, independently of the $LP_{11}$ and $LP_{21}$ combination, the HOM contribution to modal content remains less than 30 dB over the whole scanned area. Finally, it is noteworthy that such range in the FWHM/FHWM$_0$ and $\Delta d$/FHWM$_0$ variation was not increased when the fiber was coiled and shaken during the recording.

## 3. Discussion

Results reported herein demonstrate that the association of two IC claddings has a significant impact on reducing CL and modal content in HCPCFs. Specifically, we used an inner cladding made-up with a 6 suspended tube ring to ensure single-mode operation by resonantly filtering the lowest loss HOM, and outer cladding made-up with a Kagome lattice for CL enhancement. The results show that, when we surround the inner tubes by a Kagome lattice without touching them, a drop in CL by 5 orders of magnitude relative to a typical tubular or Kagome cladding HCPCF, and to the most representative reported HCPCF alternative designs can be achieved (a minimum of 8.6 × 10$^{-6}$ dB/km was obtained with the explored fiber parameters). The results give insight on how to improve the confinement in IC-HCPCF claddings, which relies on the trade-off between increasing the cladding layers (for confining light via an extremely weak power-coupling between the core and the cladding modes) and keeping their number below a critical value above-which the density of cladding modes becomes large enough to offset the cladding confining power. By making use of this concept, we developed and fabricated a hybrid cladding IC-HCPCF whereby the inner and outer cladding layers are connected via smaller and thinner tubes to keep the fiber cladding modal spectrum as close as possible to the ideal design cladding. The experimental characterization of the fiber showed a minimum loss of 1.6 dB/km at 1050nm, which is the lowest loss figure obtained at ~1 μm with HCPCFs, and a maximum extinction ratio of - 47.0 dB for the higher-order modes measured in an $S^2$ measurement using a 10 m long fiber, which the highest extinction ratio reported in HCPCFs. We found that the single-mode operation of the reported fiber is very resilient against fiber motion and laser beam misalignment. This property is particularly important for high power laser beam delivery in laser micromachining applications.

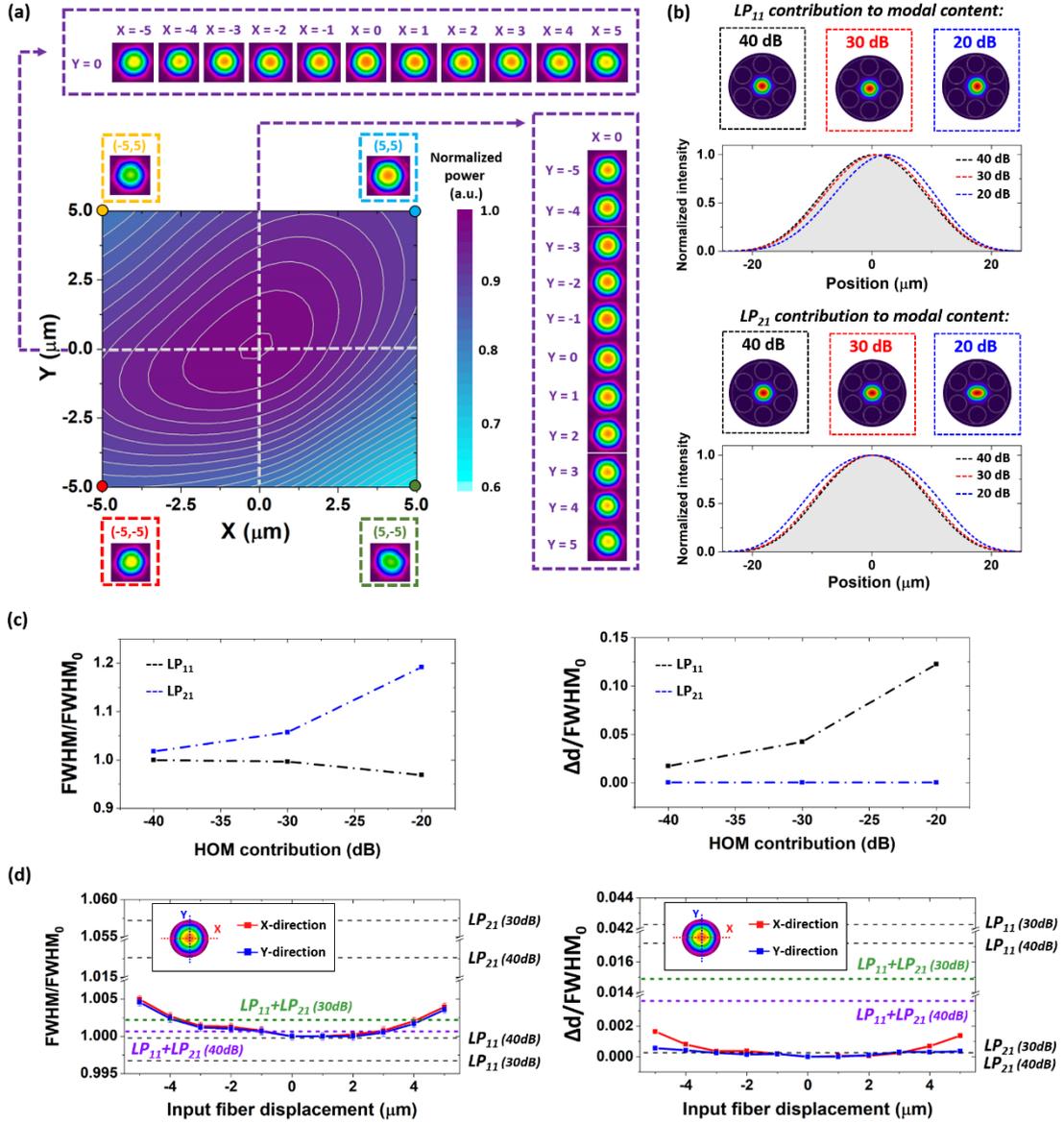

**Fig. 7.** (a) Normalized power and near field profiles at the fiber output for different input conditions at 1064 nm. (b) Intensity mode profiles considering a fiber modal content dominated by the fundamental mode and with different HOM levels added to the modal content. (c) Simulated normalized beam full width at half maximum (FWHM/FHWM$_0$) and centroid variation ($\Delta$d/FHWM$_0$) for different HOM contributions to the fiber modal content. (d) Measured FWHM/FHWM$_0$ and $\Delta$d/FHWM$_0$ of the output beam as the input fiber position was scanned along the horizontal direction, i.e., Y = 0 in (a). Horizontal lines stand for the simulated FWHM/FHWM$_0$ and $\Delta$d/FHWM$_0$ values. In the superposition cases, we assume equally weighted combinations of the LP$_{11}$ and LP$_{21}$ modes contributing to the modal content at 30 dB and 40 dB extents.

Improvements on the current HKT-HCPCF to achieve lower transmission should consider having better control of the shape and sizes of the connecting tubes between tubular and Kagome lattices during the fiber draw. By obtaining better control on this aspect, a transmission loss reduction by more than one order of magnitude is expected. Moreover, further modifications in the fiber design, such as replacing some of the six tubes with others exhibiting different confining power, will allow exploring its potential as a polarization-maintaining waveguide. The present results will contribute to the word-wide endeavors in exploring IC-HCPCFs as candidates for the next-generation long-haul optical fiber and to deepen our knowledge on the guidance mechanisms in such fibers.

## 4. Materials and methods

**Fiber fabrication:** The HKT-HCPCF was obtained by using the stack-and-draw technique in a two-step process. The first step comprehends the preform assembly and canes drawing. The second step encompasses the drawing canes to the fiber dimensions. During the fiber drawing procedure, independent pressurization was applied to the Kagome lattice, tubular lattice, and core regions to attain the required geometrical sizes.

**Cutback measurement:** In the cutback measurements, light from a supercontinuum light source was coupled to the fiber, and an optical spectrum analyzer measured the transmitted signal. For each fiber length (120 m and 4 m), the transmitted spectrum was measured for three independent fiber cleaves.

**$S^2$ measurement:** The $S^2$ measurement setup encompasses a tunable laser with a wavelength range between 1030 nm and 1070 nm (10 pm resolution) and a camera with its image acquisition routine controlled by a computer. The mode profiles and the corresponding multi-path interference values are calculated from the fiber output images acquired during laser wavelength sweeping[21]. Before the $S^2$ measurements, the fiber transmission is optimized to obtain the maximum transmitted power. MPI values shown in Fig. 6b were obtained from $S^2$ measurements taken by offsetting the fiber input from its optimum position.

**PER measurement:** In the PER measurements, light from a laser emitting at 1030 nm was launched in the fiber core, and the optical power emerging from the ports of a polarization beam splitter (PBS) was measured.

**Data availability:** The data that support the findings of this study are available from the corresponding author upon reasonable request.

**Acknowledgment.** This work is funded through PIA 4F project and the region of Nouvelle Aquitaine.

**Conflict of interests:** The authors declare that they have no conflict of interest.

**Contributions:** F.B. incepted and directed the work. F.A., J.H.O., F.D., B.D., and F.Ge. worked on fiber fabrication. J.H.O. and F.A performed the fiber characterization measurements. J.H.O. and F.B. wrote the paper. F.Gi. and L.V. performed the simulations. All the authors discussed the results and reviewed the manuscript.